\documentclass[preprint,showpacs,preprintnumbers,amsmath,amssymb,showkeys,aps,prd,a4paper,byrevtex]{revtex4}
\usepackage{graphicx}% Include figure files
\usepackage{dcolumn}% Align table columns on decimal point
\usepackage{bm}% bold math
\usepackage{psfig}
\begin{document}

\title{Meson Form Factors and Non-Perturbative Gluon Propagators}

\author{M. B. Gay Ducati}
 \email{gay@if.ufrgs.br}
\author{W. K. Sauter}
 \email{sauter@if.ufrgs.br}
\affiliation{%
Instituto de F\'{\i}sica, Universidade Federal do Rio Grande do Sul, Caixa Postal 15051, CEP 91501-970, Porto Alegre, RS, Brasil
}%

\date{\today}
             
\begin{abstract}
The meson (pion and kaon) form factor is calculated in the perturbative framework with alternative forms for the running coupling constant and the gluon propagator in the infrared kinematic region. These modified forms are employed to test the sensibility of the meson form factor to the nonperturbative contributions. Its is a powerful discriminating quantity and the results obtained with a particular choice of modified running coupling constant and gluon propagator have a good agreement with the available data, for both mesons, indicating the robustness of the method of calculation. Nevertheless, nonperturbative aspects may be included in the perturbative framework of calculation of exclusive processes.
\end{abstract}

\maketitle

\section{\label{sec:intr}Introduction}

The study of the infrared (IR) limit of the constant coupling of the Quantum Chromodynamics (QCD) had attracted much attention in the last years, as well as the infrared finite form of the gluon propagator. The behavior of the running coupling constant claimed for several theoretical and experimental studies indicates a frozen value in this kinematical region \cite{amn02}. The question of the infrared form of Green functions of QCD is still a controversial field of research (for a review see \cite{avs01,rwja}). Due to the different methods employed to analyze the Green functions, for example, the Dyson-Schwinger equations (DSE) and simulations in the lattice field theory (LFT) (as well as the approximations used to avoid several difficulties found in these methods) we have distinct different forms of Green functions. However, as we will show, the combination of the results from LFT and solutions of DSE restricts the form of the gluon propagator.

In a recent paper~\cite{amn02}, Aguilar, Mihara and Natale perform a study of the running coupling constant in the infrared region, using the dependence of the meson form factors on the coupling constant. To calculate the pion form factor, a mixed framework between perturbative and non-perturbative physics is employed with the following assumptions: freezing of the coupling constant and finite gluon propagator in the IR region, although using the perturbative scheme of calculation of the meson form factor~\cite{lb79}. 

Nevertheless, the use of a frozen running coupling constant to describe QCD exclusive processes, particularly the meson form factor, is not a novelty. Ji and Amiri~\cite{ja90} and Brodsky {\it et al.}~\cite{bjpr98} use a modified coupling constant to study the form factor and the related reaction $\gamma\gamma\rightarrow\pi^+\pi^-$. This modified coupling constant is based on the work by Cornwall~\cite{jmc82} in which a gauge independent set of Dyson-Schwinger equations is solved, giving a coupling constant finite in the infrared and a gluon propagator with a running mass. 

However, there are another approaches to include infrared contributions. In \cite{mrt}, Maris and collaborators perform a full non-perturbative calculation of the meson form factor. The amplitude for the vertex $\pi \pi \gamma$ (or $K K \gamma$), which is related with the form factor, is obtained using the following ingredients: the Bethe-Salpeter amplitudes for the scattering kernel $q \bar{q}$ (solutions of the Bethe-Salpeter equation); the solution of the DSE for the quark propagator in the rainbow truncation and an ansatz for the quark-photon vertex based in a Bethe-Salpeter equation in the ladder truncation. This approach gives a good description of the experimental data for the low momentum region for both form factors (pion and kaon) as well as another observables (see last paper of \cite{mrt}). In another work~\cite{{kcj01}}, the authors employed a light-front Bethe-Salpeter model, where the quark, instead of the gluon as the model above, has a dynamical mass, giving a good agreement with the experimental data for the pion form factor. 

The meson form factor in turn, has many attempts of description in pure perturbative QCD~\cite{vlhp}. For example, Stefanis~\cite{ngs} uses an unified factorization scheme that includes logarithmic corrections (which origin is the gluon emission) and power correction whose origin is non-perturbative. Meli\'{c} {\it et al.}~\cite{mnp99} calculate the pion form factor in next-to-leading order (NLO) in perturbation theory. Yeh~\cite{yeh02} also made a NLO calculation and found a fittable expression for the pion form factor which includes non-perturbative effects. However, these attempts have only validity where the perturbation theory is valid, in other words, when the momentum transfer is large.

In this article, we extend the results of \cite{amn02} for the pion form factor for a different set of coupling constants and propagators and compare them with more recent experimental data. Also the robustness of the model is tested with the kaon, a more massive meson than the pion. As we will show, the model employed describes both, pion and kaon form factors with good agreement with the available data for the kinematical region of exchange of low momenta. The article is organized as follows: in the section \ref{sec:mffc}, we will review the model used to calculate the meson form factor. Next, in section \ref{sec:mkff}, we present the modified meson form factor calculation scheme with the main features of the non-perturbative gluons and the frozen running coupling constant and their relation. Finally, we present the results and conclusions of the work.

%%%%%%%%%%%%%%%%%%%%%%%%%%%%%%%%%%%%%%%%%%%%%%%%%%%%%%%%%%%%%%%%%%%%%%%%%%%%%%%
%%%%%%%%%%%%%%%%%%%%%%%%%%%%%%%%%%%%%%%%%%%%%%%%%%%%%%%%%%%%%%%%%%%%%%%%%%%%%%%

\section{The Meson Form Factor Calculation \label{sec:mffc}}

\begin{figure}[ht]
\begin{center} \scalebox{1.0}{\includegraphics*[235pt,725pt][380pt,800pt]{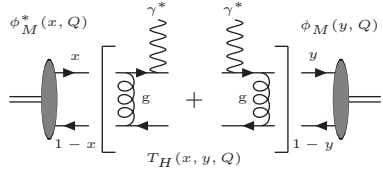}} 
\end{center}
\caption{The LO Feynman diagram for the meson form factor, where $\phi_{M}(x,\tilde{Q}^2)$ is the meson wave function.}\label{fig:fig1}
\end{figure}

The meson form factor is given by the factorized expression in perturbative QCD (pQCD) form~\cite{lb79,blk83,fld}(see figure \ref{fig:fig1})
\begin{equation}
F_{M}(Q^2)=\int^{1}_{0}\!\!dx\!\int^{1}_{0}\!\!dy\phi_M^{*}(y,\tilde{Q}_y)T_H(x,y,Q^2)\phi_M(x,\tilde{Q}_x), \label{eq:mff}
\end{equation}
where $\tilde{Q}_z={\rm min}(z,1-z)Q$ ($z=x,y$) and $Q$ is the 4-momentum transfer by the photon. The equation above can be seen as a convolution of the initial and final states, represented by the quark amplitude distributions $\phi_M(z,\tilde{Q}_z)$ (obtained in a non-perturbative calculation) with a hard (perturbative) scattering amplitude, $T_H(x,y,Q^2)$. The quark amplitude distribution is interpreted as the amplitude to find the quark or antiquark within the meson with fractional momentum $z$ or $1-z$, respectively.

The function $T_H(x,y,Q^2)$ was calculated in leading order (LO) in \cite{bt87} and in next leading order (NLO) in \cite{mnp99}. The LO expression is given by (fig. \ref{fig:fig1}):
\begin{eqnarray}
T_H(x,y,Q^2)&=& \frac{64\pi}{3Q^2}\left\{\frac{2}{3}\frac{\alpha_s[(1-x)(1-y)Q^2]}{(1-x)(1-y)} \right. \nonumber \\
&& +\left.\frac{1}{3}\frac{\alpha_s(xyQ^2)}{xy}\right\}, \label{eq:thp}
\end{eqnarray}
where $\alpha_s(Q^2)$ is the running coupling constant. 

The quark amplitude distributions $\phi_{M}(x,\tilde{Q}^2)$ are usually calculated in a certain energy scale ($Q_0$) by QCD sum rules~\cite{cz84} and then evaluated in an arbitrary energy scale solving a Bethe-Salpeter-type evolution equation~\cite{lb79,fld}, which solution is 
\begin{equation}
\phi_M(x,Q^2)= x(1-x)\sum_{n=0}^{\infty}C_{n}^{(3/2)}(2x-1)\left[ \frac{\alpha_s(Q^2)}{\alpha_s(Q_0^2)}\right]^{d_n}\phi^{(M)}_n(Q^2_0), \label{eq:mwf}
\end{equation}
where $C_{n}^{(3/2)}(2x-1)$ are the Gegenbauer orthogonal polynomials~\cite{ast};
\begin{equation}
d_n=-\frac{2A^{\rm NS}_n}{\beta_0}\,,
\end{equation}
where $A^{\rm NS}_n$ is the non-singlet anomalous dimension given by
\begin{equation}
A^{\rm NS}_n=\frac{4}{3}\left[-\frac{1}{2}+\frac{1}{(n+1)(n+2)}-2\sum_{j=2}^{n+1}\frac{1}{j}\right]\,,
\end{equation}
and $\beta_0=11-\frac{2}{3}n_f$ ($n_f$ is the number of flavors). 
To obtain the factors $\phi^{(M)}_n(Q^2_0)$, given by
\begin{equation}
\phi^{(M)}_n(Q^2_0)=\frac{4(2n+3)}{(n+1)(n+2)}\int_0^1\,dx\,C_{n}^{(3/2)}(2x-1)\phi_M(x,Q_0^2),
\end{equation}
we use the quark amplitude distributions in a given scale of energy. From \cite{cz84}, we have at $Q_0 = 500\,{\rm MeV}$ for the pion and the kaon, respectively,
\begin{subequations}
\begin{equation}
\phi_{\pi}(x,Q_0^2)=\frac{30f_{\pi}}{2\sqrt{3}}x(1-x)(2x-1)^2, \label{eq:q0fip}
\end{equation}
\begin{equation}
\phi_K(x,Q_0^2)=\frac{30f_{K}}{2\sqrt{3}}\left[0.6(2x-1)^2+0.25(2x-1)^3+0.08\right], \label{eq:q0fik}
\end{equation}
\end{subequations}
where the normalization factor $f_M$ is given by 
\begin{equation}
\int_0^1dx\;\phi_M(x,Q_0^2) = \frac{f_M}{2\sqrt{3}}, \label{eq:nrmc}
\end{equation}
and thereby $f_{\pi}=93\;{\rm MeV}$, $f_{K}=112\;{\rm MeV}$.

Using the orthogonality relation of the Gegenbauer polynomials, it is easy to find that the only non zero $\phi^{(M)}_n(Q^2_0)$'s for the pion are
\[ \phi^{(\pi)}_0(Q^2_0)=6\frac{f_{\pi}}{2\sqrt{3}},\;\; \phi^{(\pi)}_2(Q^2_0)=4\frac{f_{\pi}}{2\sqrt{3}}, \]
and for the kaon
\begin{subequations} \label{eq:phin0}
\begin{eqnarray}
\phi^{(K)}_0(Q^2_0)=6\frac{f_{K}}{2\sqrt{3}}&,&\;\; \phi^{(K)}_1(Q^2_0)=\frac{30}{28}\frac{f_{K}}{2\sqrt{3}}\;, \\
\phi^{(K)}_2(Q^2_0)=\frac{12}{5}\frac{f_{K}}{2\sqrt{3}}&,&\;\; \phi^{(K)}_3(Q^2_0)=\frac{3}{7}\frac{f_{K}}{2\sqrt{3}}\;, 
\end{eqnarray}
\end{subequations}
therefore the exponents $d_n$ are
\begin{subequations} \label{eq:dn}
\begin{eqnarray}
d_0=0&,&\;\;d_1=\frac{32}{99-6n_f}, \\
d_2=\frac{50}{99-6n_f}&,&\;\;d_3=\frac{314}{495-30n_f}.
\end{eqnarray}
\end{subequations}

Finally, the quark amplitude distribution is, for the pion,
\begin{equation}
\phi_{\pi}(x,Q^2)=\frac{f_{\pi}}{2\sqrt{3}}x(1-x)\left\{6+\left[30(2x-1)^2-6\right]\left(\frac{\alpha_s(Q^2)}{\alpha_s(Q_0^2)}\right)^{d_2}\right\}, \label{eq:pqad}
\end{equation}
and for the kaon,
\begin{eqnarray}
&&\phi_K(x,Q^2)=\frac{f_K}{2\sqrt{3}}x(1-x) \nonumber \\ 
&&\times \left\{6+\frac{45}{14}(2x-1)\left(\frac{\alpha_s(Q^2)}{\alpha_s(Q^2_0)}\right)^{d_1}+ \nonumber \right. \\ 
&&\left.+\frac{6}{5}\left[15(2x-1)^2-3\right]\left(\frac{\alpha_s(Q^2)}{\alpha_s(Q^2_0)}\right)^{d_2}+ \right. \nonumber \\
&&\left.+\frac{3}{14}\left[35(2x-1)^3-15(2x-1)\right]\left(\frac{\alpha_s(Q^2)}{\alpha_s(Q^2_0)}\right)^{d_3}\right\}. \label{eq:kwf}
\end{eqnarray}

However, Dziembowski and Mankiewicz~\cite{dm87} use the constituent-quark model to calculate the quark amplitude distribution for the pion, obtaining
\begin{equation}
\phi^{({\rm DM})}_{\pi}(x)=N\exp\left[-\frac{m^2}{8x(1-x)\beta^2}\right]\left\{\frac{(xM+m)\left[(1-x)M+m\right]}{4\beta^4}-2x(1-x)\right\}, \label{eq:qaddm}
\end{equation}
where $N$ is determined by the normalization condition Eq.(\ref{eq:nrmc})($N=0.622$), $M$ is the spin-averaged meson mass ($M=614.4$ MeV for the pion), $m$ is the constituent quark mass ($m=330$ MeV) and $\beta$ is a Gaussian parameter chosen as $460$ MeV. We will test the changes in the pion form factor when this quark distribution is employed. 

%%%%%%%%%%%%%%%%%%%%%%%%%%%%%%%%%%%%%%%%%%%%%%%%%%%%%%%%%%%%%%%%%%%%%%%%%%%%%%%
%%%%%%%%%%%%%%%%%%%%%%%%%%%%%%%%%%%%%%%%%%%%%%%%%%%%%%%%%%%%%%%%%%%%%%%%%%%%%%%
\section{The modified meson form factor} \label{sec:mkff}

In \cite{amn02}, the modifications introduced in the pion form factor by the change of the usual running coupling amplitude by the frozen infrared one have been tested, as well as the case of the change of the gluon propagator. In this case, the hard scattering amplitude can be written as
\begin{equation}
{\tilde{T}}_H(x,y,Q^2)= \frac{64\pi}{3}\left\{\frac{2}{3}{\tilde{\alpha}}_s(\hat{k}^2)D(\hat{k}^2)+\frac{1}{3}{\tilde{\alpha}}_s(\hat{p}^2)D(\hat{p}^2)\right\}, \label{eq:thpm}
\end{equation}
where $\hat{k}^2=(1-x)(1-y)Q^2$, $\hat{p}^2=xyQ^2$ and $D(Q^2)$ is the gluon propagator, and  ${\tilde{\alpha}}_s$ is the modified running coupling constant which expression is given below.

Therefore, the expression for the meson form factor from Eq.(\ref{eq:mff}) and (\ref{eq:thpm}) is
\begin{equation}
{\tilde{F}}_{M}(Q^2)=\int^{1}_{0}\!\!dx\!\int^{1}_{0}\!\!dy\,\tilde{\phi}_M^{*}(y,\tilde{Q}_y){\tilde{T}}_H(x,y,Q^2)\tilde{\phi}_M(x,\tilde{Q}_x), \label{eq:mkff}
\end{equation}
where $\tilde{\phi}_M$ is the meson wave function with the modified running coupling constant. 

The infrared form of the gluon propagator is still a controversial aspect (for a review for solutions and methods see \cite{rwja} and references therein). The general formula for the gluon propagator is given by (in the Landau gauge):
\begin{equation}
D^{ab}_{\mu\nu}(q^2)=\delta^{ab}\left(\delta_{\mu\nu}-\frac{q_{\mu}q_{\nu}}{q^2}\right)D(q^2), \label{eq:gp}
\end{equation}
where the usual perturbative propagator is given by $D(q^2)=1/q^2$ which diverges when $q^2 \rightarrow 0$. In the literature, there are several different forms for the gluon propagator, with different behaviors for the infrared region: finite, zero and more divergent that $1/k^2$, each one with its advantages and inconveniences. The reasons of these different behaviors are the methods and approximations employed to obtain the gluon propagator. The more popular methods are the Schwinger-Dyson equations and the simulations from lattice field theory. The most recent results in the last method~\cite{fdrb01,lrg02} discard the solution of the type $1/k^4$ and therefore we will not use it in this work. In spite of the controversy, the non-perturbative gluon propagator was used successfully in many phenomenological applications as, for example, proton-proton scattering~\cite{hkpr} and elastic production of vector mesons~\cite{ghnw}.

In this work, we use the following gluon propagators:
\begin{itemize}
\item calculated by Cornwall~\cite{jmc82} using the gauge independent pinch technique, given by
\begin{equation}
D_{\rm C}(q^2)=\frac{1}{q^2+M^2_g(q^2)}, \label{eq:cgp}
\end{equation}
where $M^2_g(q^2)$ is a dynamical gluon mass term,
\begin{equation}
M^2_g(q^2)=m_g^2\left\{\frac{\ln\left(\frac{q^2+4m^2_g}{\Lambda_{\rm QCD}^2}\right)}{\ln\left(\frac{4m^2_g}{\Lambda_{\rm QCD}^2}\right)}\right\}^{-12/11} \label{eq:mgt}
\end{equation}
with $m_g^2=500\pm200\;{\rm MeV}$. With this propagator, \cite{amn02} obtains the best description for the pion form factor. Its features include the correct ultraviolet behavior (according to the renormalization group) and the dynamically generated mass $M^2_g(q^2)$. 

\item from H\"abel {\it et al.}~\cite{hkrsw90}, calculated using an approach that employs the same features of the perturbative theory, which implies a simplified set of Schwinger-Dyson equations, which solution is given by
\begin{equation}
D_{\rm H}(q^2)=\frac{1}{q^2+\frac{b^2}{q^2}}, \label{eq:hgp}
\end{equation}
We should note that this propagator goes to zero when $q^2 \rightarrow 0$, unlike the other propagators employed in this work, which have a non-zero value in this region. 

\item Alkofer {\it et al.}~\cite{ahs} solve a set of Dyson-Schwinger equations for the QCD Green's functions and the gluon propagator found by this manner can be fitted~\cite{amn02} by 
\begin{equation}
D_{\rm A}(q^2)=\frac{bq^2}{q^4+c^2}, \label{eq:agp}
\end{equation} 
where
$b=3.707$ and $c=0.603$.

\item Atkinson and Bloch~\cite{abw} solve a set of Dyson-Schwinger equations for the gluon and ghost propagator, improving the solution found by \cite{ahs}. The asymptotic infrared behavior of the gluon propagator is
\begin{equation}
D_{\rm AB}^{\rm IR}(q^2) \sim \frac{1}{q^2}\left\{A_0(q^2)^{2\kappa}\left(1+\sum_{\lambda=1}^3 f_{\lambda}a_{\lambda}(q^2)^{\lambda \rho}\right)\right\} \label{eq:abgp},
\end{equation}
where $A_0=1$, $\kappa=0.769475$, $a_1=-10.27685$, $\rho=1.96964$, $f_1=1$, $f_2=0.408732$, $f_3=-0.761655$.

\item from Gorbar and Natale~\cite{gn00}, that use the operator product expansion (OPE) to relate the gluon and quark propagators with their respective condensates, the gluon propagator obtained is 
\begin{equation}
D_{\rm GN}(q^2)=\frac{1}{q^2+\mu_g^2\,\theta(\chi\mu_g^2-q^2)+\frac{\mu_g^4}{q^2}\,\theta(q^2-\chi\mu_g^2)}, \label{eq:gngp}
\end{equation}
where $\mu^2_g = \,0.61149\,{\rm GeV}$ is a parameter fixed by the gluon condensate~\cite{gn00} and $\chi=$0.9666797. We should note that this propagator interpolates two different propagators: the pure massive and one similar to the H\"abel solution.

\end{itemize}

%\subsection{The running coupling constant} \label{sec:rcc}
The modification of the gluon propagator is closely related with the modification of the running coupling constant in the infrared region. As calculated by Cornwall~\cite{jmc82}, the coupling constant is frozen in the low momentum by the addition of the massive term, Eq.(\ref{eq:mgt}),
\begin{equation}
\alpha_s^{(\rm C)}(q^2)=\frac{4\pi}{\beta_0\ln\left(\frac{q^2+4M^2_g(q^2)}{\Lambda^2_{\rm QCD}}\right)}, \label{eq:r3c}
\end{equation}
where $\Lambda^2_{\rm QCD}$ is the QCD scale parameter and $\beta_0=11-\frac{2}{3}n_f$ is the first coefficient of the beta function and $n_f$ is the number of flavors. In the case of the other propagators, we consider the terms in the denominator as massive terms and we will substitute it in the Eq.(\ref{eq:r3c}) as well as in the Cornwall propagator, as shown below
\begin{itemize}
\item The H\"abel propagator gets
\begin{equation}
\alpha_s^{(\rm H)}(q^2)=\frac{4\pi}{\beta_0\ln\left(\frac{q^2+M^2_{\rm H}(q^2)}{\Lambda^2_{\rm QCD}}\right)}, \label{eq:rcch}
\end{equation}
with
\[ M^2_{\rm H}(q^2)= \frac{b^2}{q^2}. \]
\item The Gorbar-Natale propagator gets
\begin{equation}
\alpha_s^{(\rm GN)}(q^2)=\frac{4\pi}{\beta_0\ln\left(\frac{q^2+M^2_{\rm GN}(q^2)}{\Lambda^2_{\rm QCD}}\right)}, \label{eq:rcgn}
\end{equation}
with
\[M^2_{\rm GN}(q^2)=\mu_g^2\,\theta(\chi\mu_g^2-q^2)+\frac{\mu_g^4}{q^2}\,\theta(q^2-\chi\mu_g^2).  \]
\end{itemize}

However, in the case of the propagators given by the equations (\ref{eq:agp}) and (\ref{eq:abgp}), the coupling constant comes from the solutions of the Dyson-Schwinger equations for the Green's functions. For the case of the solution of the Alkofer {\it et al.}, the coupling constant can be fitted by the following formula~\cite{amn02}
\begin{eqnarray}
\alpha_{s}^{\rm A}(q^2)=\left\{\begin{array}
{l@{\quad:\quad}l}
\alpha_{s}^{\rm a}&   q^2 < 0.31 \;  {\rm GeV}^2\\
\alpha_{s}^{\rm b}& 0.31 < q^2 < 1.3 \; {\rm GeV}^2\, , \\
\alpha_{s}^{\rm c}& q^2>1.3\;  {\rm  GeV}^2
\end{array} \right. \label{eq:asa}
\end{eqnarray}
with
\begin{eqnarray}
\alpha_{s}^{\rm a}(q^2)&=&0.2161+9.2621\exp\left(-2\frac{(q^2-0.0297)^2}{(0.6846)^2}\right), \nonumber \\ 
\alpha_{s}^{\rm b}(q^2)&=&1.4741+8.6072\exp\left(-\frac{q^2-0.1626}{0.3197}\right), \nonumber   \\
\alpha_{s}^{\rm c}(q^2)&=&  \frac{1.4978}{\ln (1.8488 q^2)} . 
\end{eqnarray}

The solution of Atkinson and Bloch~\cite{abw} is given analytically in the asymptotic regions
\begin{equation}
\alpha^{{\rm AB}}_{s}(q^2) \stackrel{\rm IR}{\sim}4\pi \nu \left[1-\sum_{k=1}^{3}b_k\left(\frac{q^2}{\Omega^2}\right)^{k \rho}\right], \label{eq:aabr}
\end{equation}
where $\nu=0.912771$, $\Omega^2=0.1864754$, $b_1=-1$, $b_2=0.760753$, $b_3=-0.370785$, and
\begin{equation}
\alpha^{{\rm AB}}_{s}(q^2) \stackrel{\rm UV}{\sim}\frac{4\pi}{4\log(\frac{q^2}{\Lambda^2_{\rm QCD}})}, \label{eq:aabu}
\end{equation} 
where $\Lambda^2_{\rm QCD}=0.06802\,{\rm GeV}^2$.

The procedure is the following: using the Eq.(\ref{eq:mkff}), we substitute in the hard scattering amplitude $\tilde{T}_H$ the different gluon propagators in the order above as well as the running coupling constant, $\tilde{\alpha}_s$, to verify the changes in the form factor. A comparison with the perturbative fit of Yeh~\cite{yeh02}, as well as, with the full non-perturbative calculation of Maris and collaborators~\cite{mrt} is also performed.

%%%%%%%%%%%%%%%%%%%%%%%%%%%%%%%%%%%%%%%%%%%%%%%%%%%%%%%%%%%%%%%%%%%%%%%%%%%%%%%
%%%%%%%%%%%%%%%%%%%%%%%%%%%%%%%%%%%%%%%%%%%%%%%%%%%%%%%%%%%%%%%%%%%%%%%%%%%%%%%
\section{Results} \label{sec:rec}

\begin{figure}[t]
\vspace{5mm}
\begin{center}
\scalebox{.65}{\includegraphics*{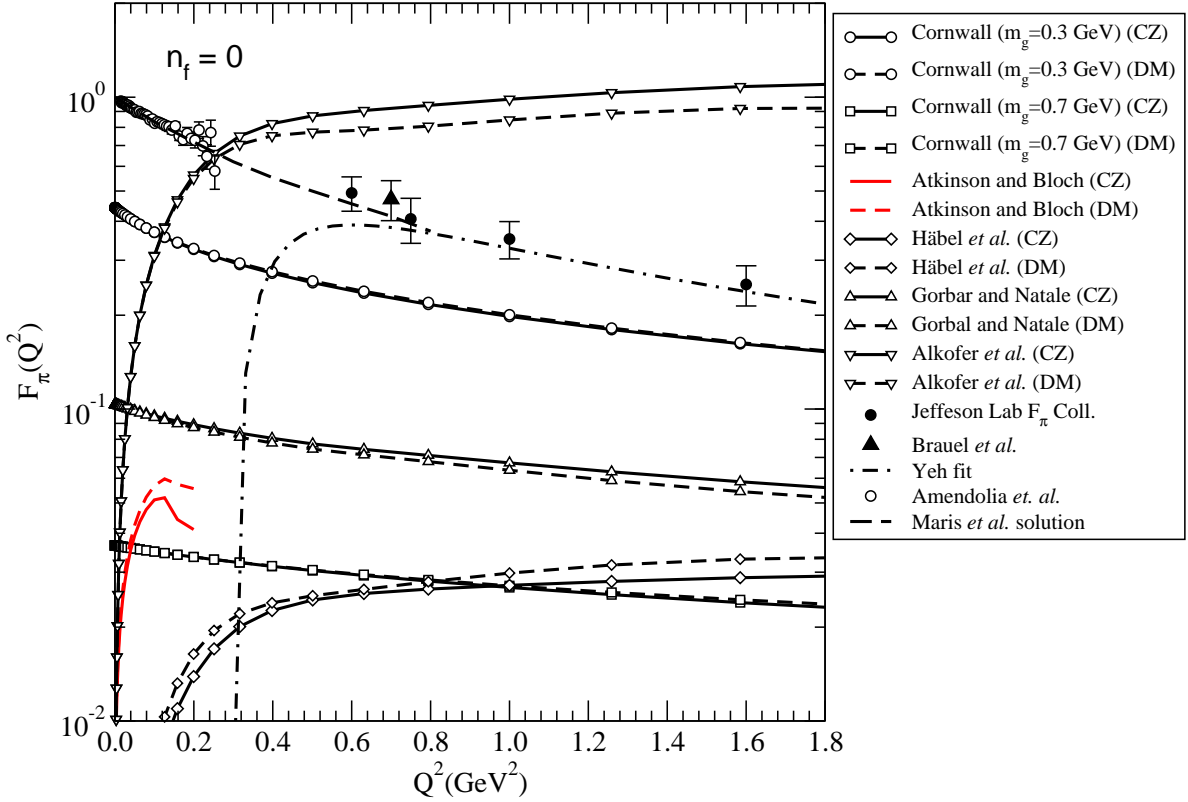}}
\end{center}
\caption{The pion form factor calculated using the modified running coupling constant and gluon propagator from different works, indicated in the table above. The number of flavors is zero. The labels CZ and DM refer to the eqs.\protect{(\ref{eq:pqad})} and \protect{(\ref{eq:qaddm})} for the quark amplitude distributions. The solution of Maris {\it et al.} from \protect{\cite{mrt}}. The Yeh fit from eq. \protect{(\ref{eq:ypf})}. Data from \protect{\cite{amen86,jflb01}}. }  \label{fig:fig2}
\end{figure}

\begin{figure}[t]
\vspace{5mm}
\begin{center}
\scalebox{.65}{\includegraphics*{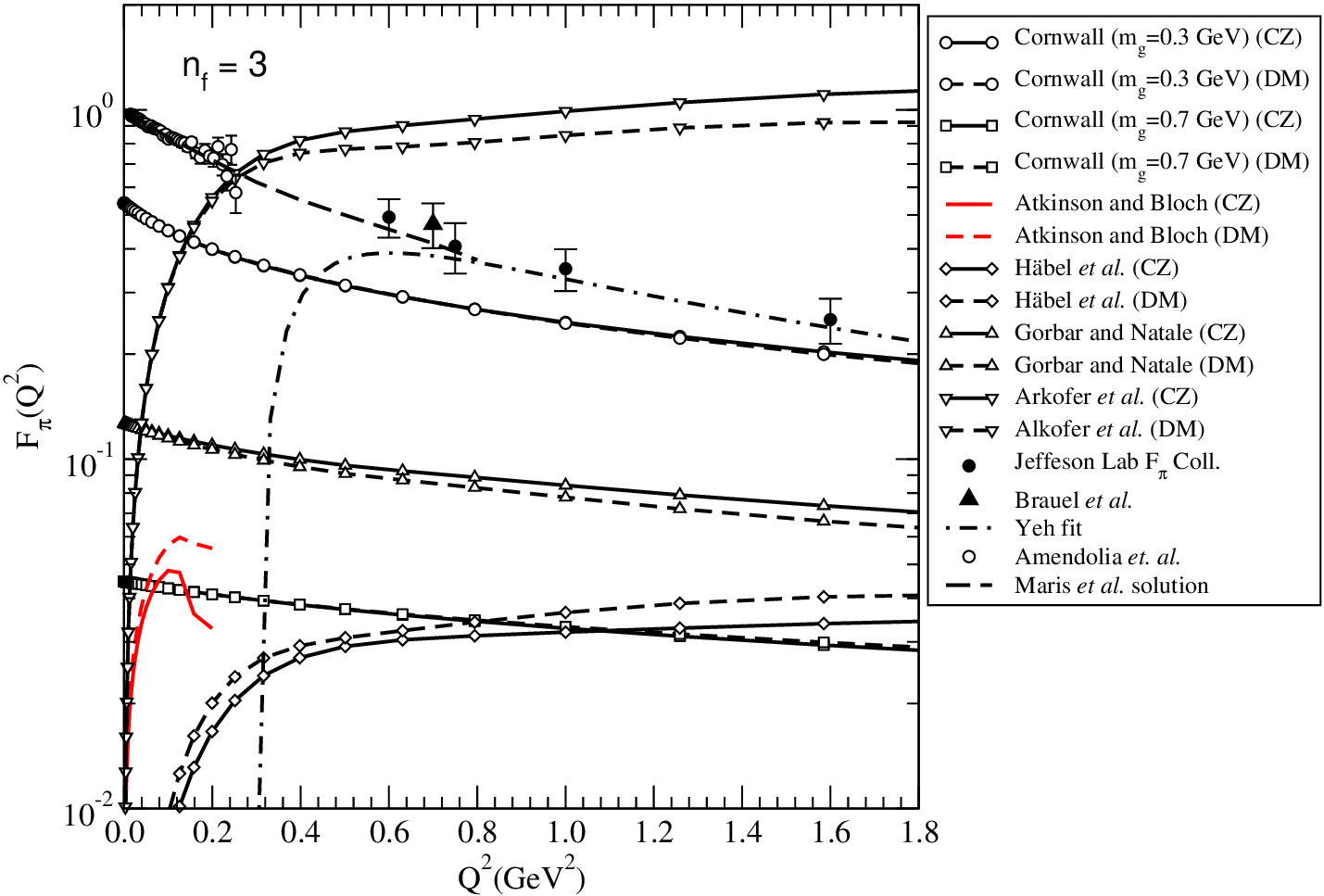}}
\end{center}
\caption{The pion form factor calculated as in figure \protect{\ref{fig:fig2}} with $n_f=3$.} \label{fig:fig3}
\end{figure}

\begin{figure}[t]
\vspace{5mm}
\begin{center}
\scalebox{.65}{\includegraphics*{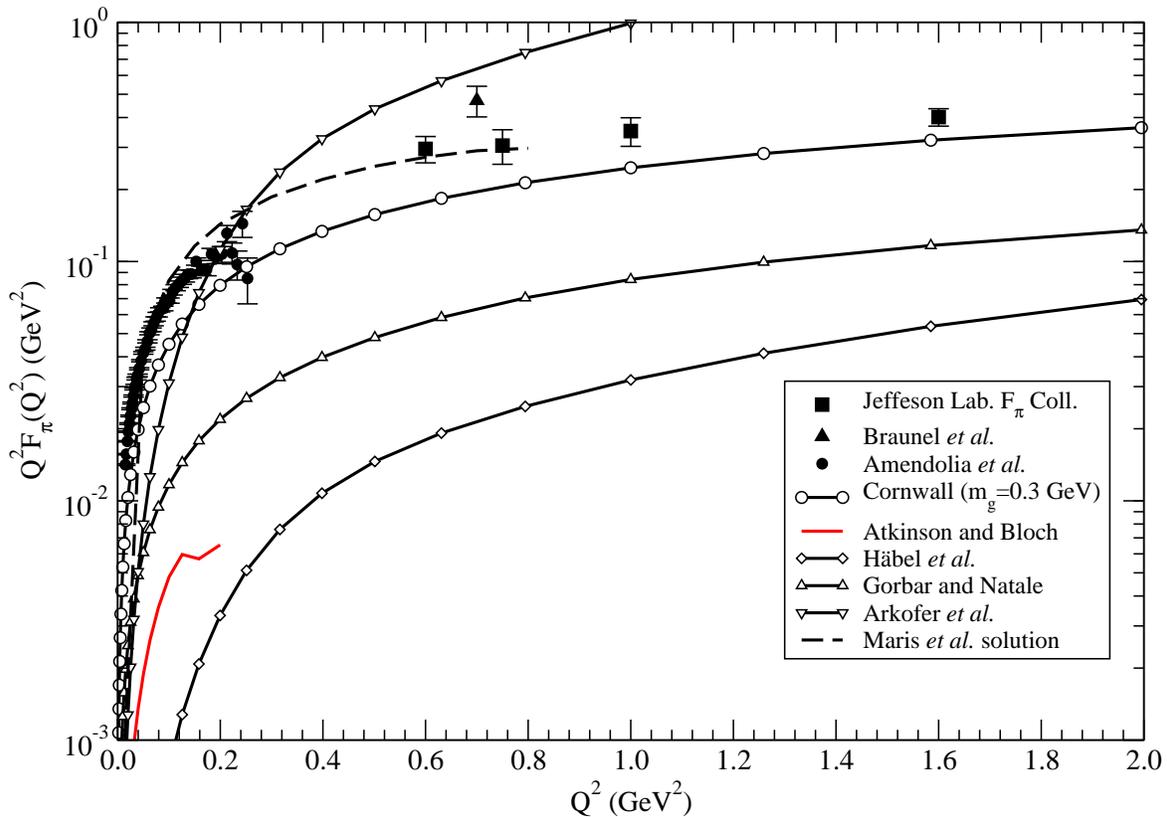}}
\end{center}
\caption{Plot of $Q^2F_{\pi}$ versus $Q^2$ using the same coupling constant and propagators used in the fig. \protect{(\ref{fig:fig2})}, with quark amplitude distribution from Eq. (\ref{eq:pqad}) and $n_f=3$.} \label{fig:qpqf}
\end{figure}

\begin{figure}[t]
\vspace{5mm}
\begin{center}
\scalebox{.65}{\includegraphics*{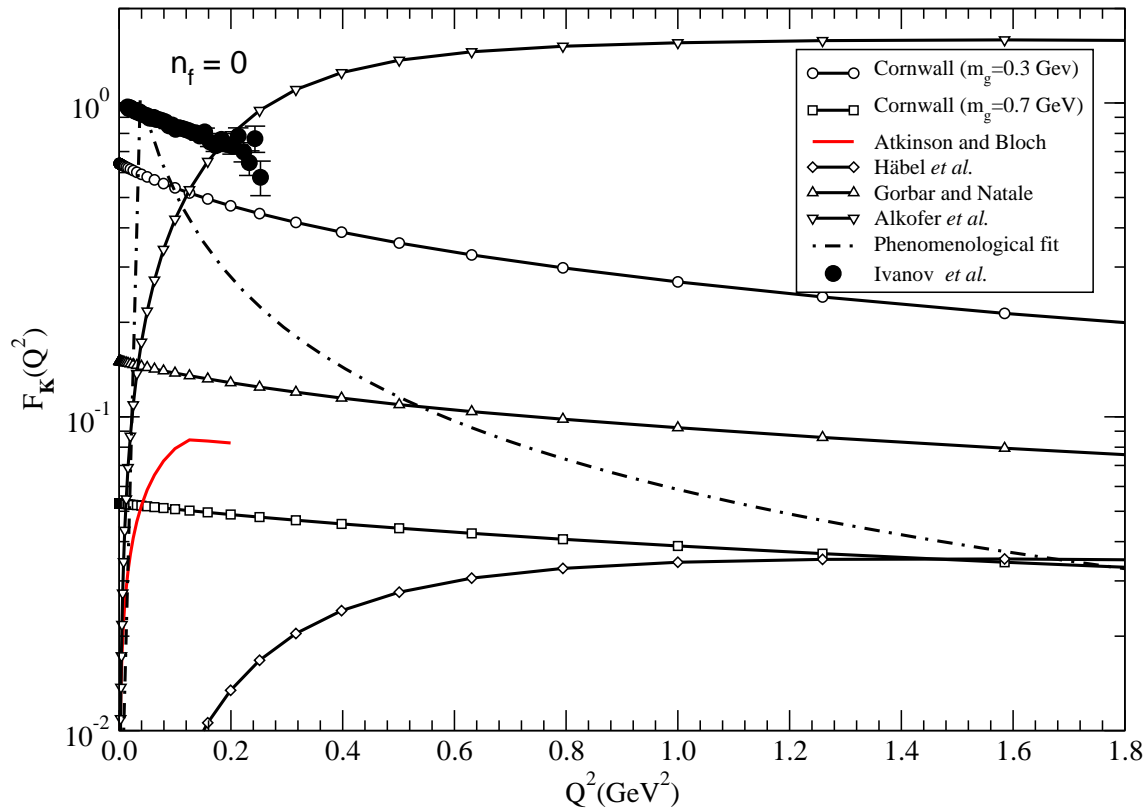}}
\end{center}
\caption{The kaon form factor calculated as the pion form factor, but using only the quark amplitude distribution from Eq. \protect{(\ref{eq:kwf})}. The phenomenological fit is given by Eq. \protect{(\ref{eq:kfit})}. Data from \protect{\cite{iv81}}} \label{fig:qfk}
\end{figure}

\begin{figure}[t]
\vspace{5mm}
\begin{center}
\scalebox{.65}{\includegraphics*{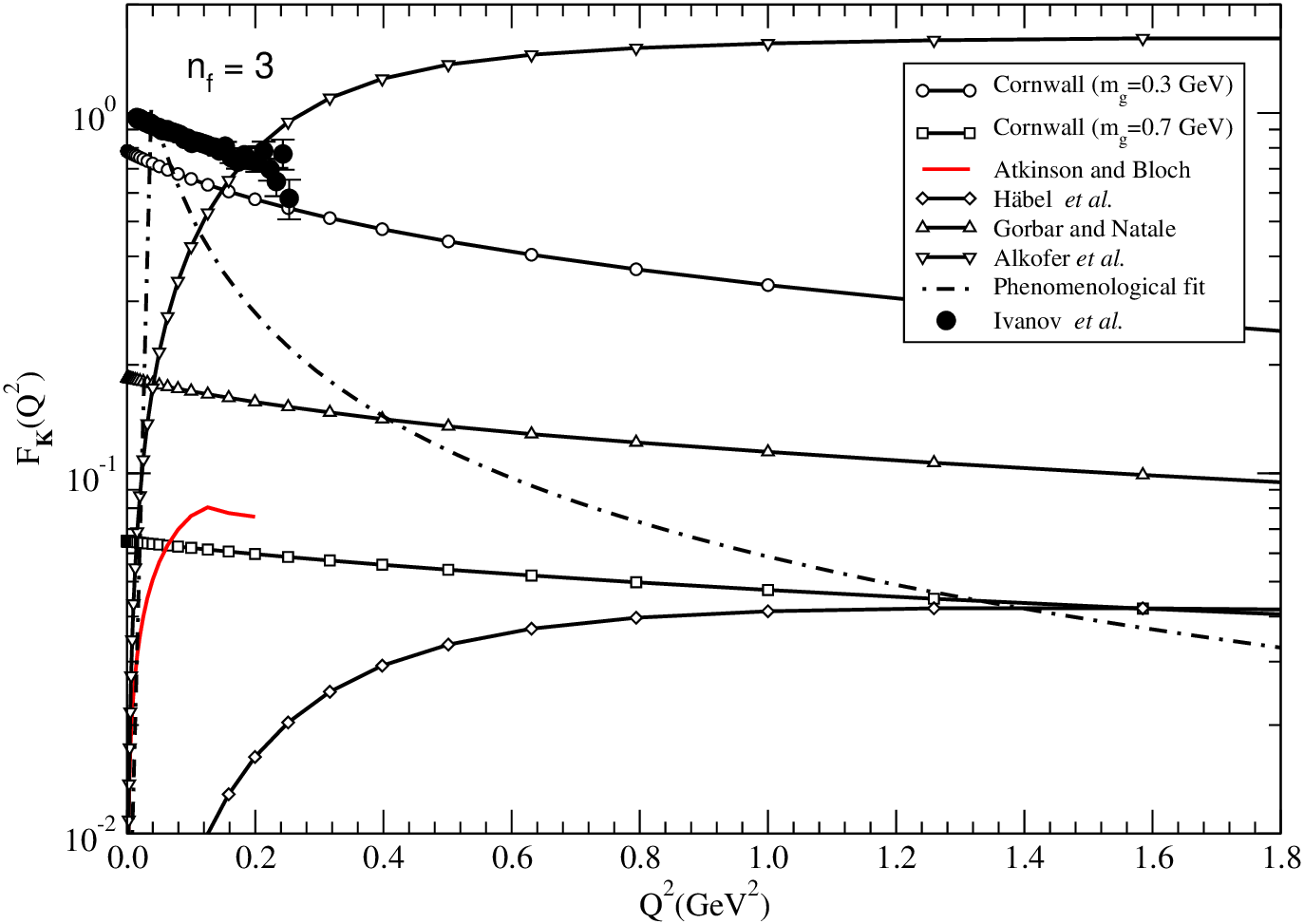}}
\end{center}
\caption{The kaon form factor calculated as in figure \protect{(\ref{fig:qfk})}, with $n_f=3$} \label{fig:qfk3}
\end{figure}

\begin{figure}[t]
\vspace{5mm}
\begin{center}
\scalebox{.65}{\includegraphics*{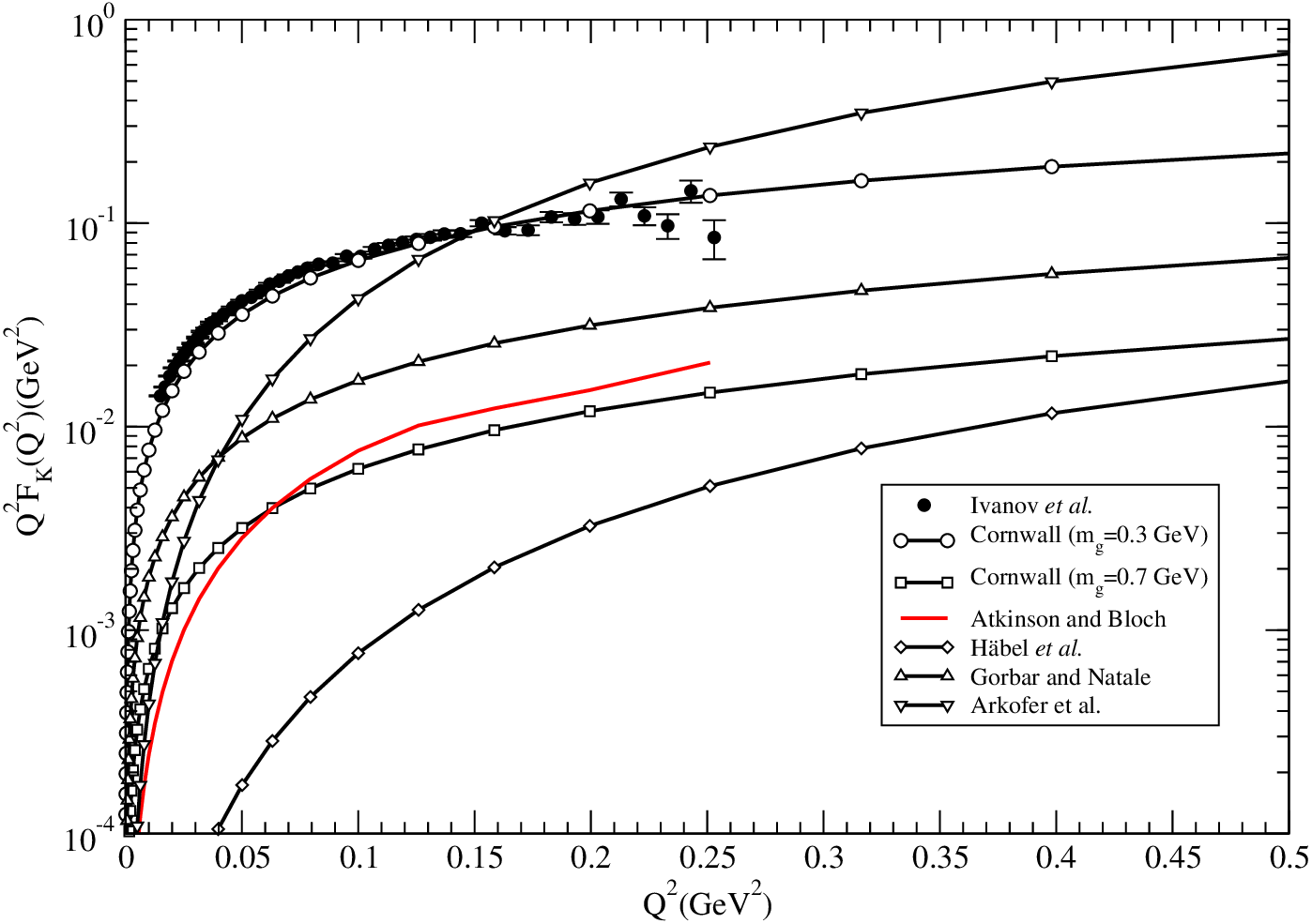}}
\end{center}
\caption{Plot of $Q^2F_{K}$ versus $Q^2$ using same coupling constant and propagators used in the fig. \protect{(\ref{fig:qfk})}, with quark amplitude distribution from Eq. (\ref{eq:kwf}) and $n_f=3$.} \label{fig:qkqf}
\end{figure}

The results obtained for the pion form factor in the flavorless case ($n_f=0$) are displayed in the figure \ref{fig:fig2}. A remarkable feature of the result is the significant difference between the distinct coupling constants and their respective gluon propagators. The origin of this difference is the strong dependence of the form factor in the running coupling constant. The results are also consistent with the infrared behavior of the coupling constant, for example, in the case of the H\"abel propagator (Eq. (\ref{eq:rcch})) when $Q^2 \rightarrow 0$ the massive term diverges, which implies a divergent behavior. With the other propagators, the behavior is distinct because the coupling constant is finite when $Q^2 \rightarrow 0$, giving a finite result. 

As already pointed out in \cite{amn02}, the best description for the data is given by the Cornwall propagator with $m_g=300\;{\rm MeV}$ while the other choices for the propagator listed above do not give a reasonable result in comparison with the available experimental data.

When a more realistic number of flavors, $n_f=3$, is used, the result for the form factor does not present a significant change, as expected in \cite{amn02} and displayed in fig. (\ref{fig:fig3}), however the result with three flavors is better than with no flavors.

As the number of flavors, the change in the quark amplitude distribution, Eq. (\ref{eq:pqad}) and (\ref{eq:qaddm}) does not make a significant modification in the form factor.

We display in the figures (\ref{fig:fig2}), (\ref{fig:fig3}) and (\ref{fig:qpqf}), the full non-perturbative solution for the pion form factor obtained by Maris and collaborators~\cite{mrt}, only in the low momentum transfer region ($Q^2 \leq 0.8\,{\rm GeV}^2 $). As pointed out in the Introduction, this solution gives a very good description of the low momentum data. Our best result, using the Cornwall's propagator and $m_g= 0.3$ GeV, is below of this prediction although with the same global behavior. If we diminish the smaller massive parameter $m_g$, we can obtain a better fit to the data (for both mesons), but this value is outside of the range of the values for this parameter, which provides good description of another processes~\cite{hkpr,ghnw}.  

We also display a fit with the experimental points, based on the approach of \cite{yeh02}. In that paper, the author uses the collinear expansion to analyze the NLO power corrections to the pion form factor, and founds the following expression for the pion form factor
\begin{equation}
F_{\pi}(Q^2)=\frac{16\pi\alpha_s(Q^2_{\rm eff})f_{\pi}^2}{Q^2}\left(1-\frac{32\pi^2f^2_{\pi}J(Q^2_{\rm eff})}{Q^2}\right), \label{eq:yeh}
\end{equation}
where $Q^2_{\rm eff}$ is the effective energy scale and $J(Q^2_{\rm eff})$ is called a jet function, introduced to absorb the infrared divergences from the NLO corrections. In the same paper, it was made a phenomenological fit to the data for the pion form factor with the following form
\begin{equation}
F^{\rm fit}_{\pi}(Q^2)=\frac{A}{Q^2}\left(1-\frac{B}{Q^2}\right), \label{eq:ypf}
\end{equation}
In \cite{yeh02}, the best values for $A$ and $B$ to fit the above formula to the experimental data from \cite{bb76,amen86} are $A=0.46895$ and $B=0.3009$.

The phenomenological fit has a divergence when the momentum transferred by the photon goes to zero, in opposition to the results that employ the non-perturbative propagators and a finite infrared running coupling constant, and therefore does not give a good description for the data in low $Q^2$.

In order to compare with the previous results found in the literature, we plot $Q^2\;F_{\pi}(Q^2)$ as a function of $Q^2$, as shown in the fig. \ref{fig:qpqf}. In \cite{ja90}, its prediction does not go to zero, as in our result, neither describes the data in the region of low $Q^2$. In \cite{jflb01}, a comparison of predictions found in the literature for the pion form factor is made and the best result of this study (Cornwall with $m_g=300$ MeV) still is a good result.

In the case of the kaon, we used the same procedure, although using only the quark amplitude distribution of the Eq.(\ref{eq:kwf}) in the Eq.(\ref{eq:thpm}). The result of the flavorless case ($n_f=0$) is displayed in the fig.(\ref{fig:qfk}) while the case $n_f=3$ is shown in the fig.(\ref{fig:qfk3}). As shown in the figures, the global behavior is the same as in the pion form factor, in other words, the best description for the data is given again by the Cornwall's choices for the coupling constant and gluon propagator with $m_g=300$ MeV.

In the figures (\ref{fig:qfk}) and (\ref{fig:qfk3}) we present a result, based in \cite{yeh02}, with the same functional form of the pion, but now applied to the available data for the kaon,
\begin{equation}
F^{\rm fit}_{K}(Q^2)=\frac{C}{Q^2}\left(1-\frac{D}{Q^2}\right), \label{eq:kfit}
\end{equation}
since the only dependence of the meson in the Eq.(\ref{eq:yeh}) is the normalization factor $f_M$. We compare Eq.(\ref{eq:kfit}) to the data available for the kaon~\cite{iv81} and found for the coefficients $C=0.0594695$ and $D=0.0130963$ with $\chi^2=1.93314$.

As the data available for the kaon restricts to the region of low $Q^2$, the predictions of different models cannot be tested in the kinematic region of high $Q^2$ for the moment, in opposition to the case of the pion.

For the pion, as well as for the kaon, when the coupling constant and the propagator of Atkinson and Bloch~\cite{abw} work is employed, due to the absence of a full analytical form for the coupling constant and for the gluon propagator, we use only the infrared analytical forms, exception to the case of the fixed scale of energy (see Eq.(\ref{eq:q0fip})), when we use the ultraviolet analytical result. There is a difference in the full numerical result and the analytical asymptotic expression for the coupling constant (see the first paper of \cite{abw}), but it is not significant for our result.

%%%%%%%%%%%%%%%%%%%%%%%%%%%%%%%%%%%%%%%%%%%%%%%%%%%%%%%%%%%%%%%%%%%%%%%%%%%%%%%%%%%%%%%%%%%%
\section{Conclusions} \label{sec:conc}

In this work, we calculated the pion and kaon form factor following the approach of \cite{amn02} obtaining a reasonable agreement with the data available (for the pion~\cite{bb76,amen86,jflb01} and kaon~\cite{iv81}) in the case of the Cornwall's gluon propagator~\cite{jmc82} (with parameter $m_g=300\;{\rm GeV}$) and the running coupling constant that is frozen in the small momentum transfer region, avoiding infrared divergences. We point out that a smaller mass gives a better result, but its value is out of the mass interval found in previous results for another processes~\cite{ghnw}. The other propagators (Cornwall with $m_g=700\;{\rm GeV}$, H\"abel {\it el al.}~\cite{hkrsw90}, Alkofer {\it el al.}~\cite{ahs}, Atkinson and Bloch~\cite{abw} and Gorbar and Natale~\cite{gn00}) give results in disagreement with the experimental results. The significant difference between the results has its origin in the well-known meson form factor sensibility to the running coupling constant. Since we modified the IR behavior of the $\alpha_s$, this difference is expected as in \cite{amn02}, where the same behavior is observed. The case with a non-zero number of flavors, $n_f=3$, does not have a significant difference in the results. The exchange of the quark amplitude distribution in the pion case also does not modify significantly the form factor according to \cite{ja90}.  The model employed also describes the available experimental data in the low $Q^2$ region for both (pion, kaon) mesons form factors, in opposition to the results~\cite{ja90} that employ only a frozen form of the running coupling constant, without the finite infrared form for the gluon propagator, but agrees with the results shown in \cite{jflb01} for the pion form factor. In \cite{mnp99,rad91}, restrictions are made concerning the applicability of the frozen coupling constant and the modified gluon propagator. In \cite{rad91} the pion form factor is calculated using the same ideas of this work, although employing a different pion wave function and a gluon propagator with fixed mass, resulting for $Q^2F_{\pi}(Q^2)$ a smaller value than the  experimental one by a factor of 10. In comparison with the pure non-perturbative calculation of \cite{mrt} and the NLO perturbative calculation of \cite{yeh02}, our results for the pion (except by a normalization factor, given by adjusting the massive parameter) is an interpolation of the above results, having the fit characteristics of both calcutations. The good agreement with the experimental data of this work indicates that it is possible to include non-perturbative effects in exclusive processes in QCD without dramatic changes in the perturbative scheme of calculation of observables. Otherwise, the inclusion of non-perturbative effects is still a field in which much research is required.   

%We also tested the model of the form factor with a different gluon propagator and a frozen running coupling constant, which remains describing the data as in the pion case.

%%%%%%%%%%%%%%%%%%%%%%%%%%%%%%%%%%%%%%%%%%%%%%%%%%%%%%%%%%%%%%%%%%%%%%%%%%%%%%%%%%%%%%%%%%%%
\begin{acknowledgments}
This research work was supported by the Conselho Nacional de Desenvolvimento Cient\'{\i}fico e Tecnol\'ogico (CNPq). The authors thank A. N. Natale for the fruitful discussions during this work.
\end{acknowledgments}

%%%%%%%%%%%%%%%%%%%%

\end{document}